# Beyond Labels: Zero-Shot Diabetic Foot Ulcer Wound Segmentation with Self-attention Diffusion Models and the Potential for Text-Guided Customization

*Abderrachid Hamrani [1] \*, Daniela Leizaola [2], Renato Sousa [3], Jose P. Ponce [3], Stanley Mathis [3][4], David G. Armstrong [5], Anuradha Godavarty [2] \**

*Abstract*—Diabetic foot ulcers (DFUs) pose a significant challenge in healthcare, requiring precise and efficient wound assessment to enhance patient outcomes. This study introduces the Attention Diffusion Zero-shot Unsupervised System (ADZUS), a novel text-guided diffusion model that performs wound segmentation without relying on labeled training data. Unlike conventional deep learning models, which require extensive annotation, ADZUS leverages zero-shot learning to dynamically adapt segmentation based on descriptive prompts, offering enhanced flexibility and adaptability in clinical applications. Experimental evaluations demonstrate that ADZUS surpasses traditional and state-of-the-art segmentation models, achieving an IoU of 86.68% and the highest precision of 94.69% on the chronic wound dataset, outperforming supervised approaches such as FUSegNet. Further validation on a custom-curated DFU dataset reinforces its robustness, with ADZUS achieving a median DSC of 75%, significantly surpassing FUSegNet's 45%. The model's text-guided segmentation capability enables real-time customization of segmentation outputs, allowing targeted analysis of wound characteristics based on clinical descriptions. Despite its competitive performance, the computational cost of diffusion-based inference and the need for potential fine-tuning remain areas for future improvement. ADZUS represents a transformative step in wound segmentation, providing a scalable, efficient, and adaptable AI-driven solution for medical imaging.

## I. INTRODUCTION

Diabetic foot ulcers (DFUs) are a prevalent and severe complication among individuals with diabetes, significantly impacting patient morbidity and healthcare costs[1], [2]. As a critical manifestation of diabetic neuropathy and peripheral vascular disease, DFUs affect approximately 34% of the diabetic population at some point in their lives[1], [3], [4]. The risk of lower extremity amputation increases substantially with the presence of ulcers, making early and accurate detection vital for effective intervention and management [5]. However, the standard procedures for detecting and assessing the severity of DFUs largely depend on physical examinations conducted by healthcare professionals [6]. This traditional approach requires significant clinical expertise and is inherently limited by its subjective nature, leading to variability in diagnosis and treatment outcomes. The reliance on manual inspection makes the process not only labor-intensive but also inconsistent, with potential disparities in care depending on the practitioner's experience and the clinical setting [7]. Visual inspections are challenged by factors such as poor lighting, diverse skin tones, and the subtle appearance of early ulcers, which may not be distinctly visible [8], [9]. As a result, there is a pressing need for more objective, reliable, and scalable methods for DFU detection and monitoring that can support clinicians [10]–[12].

Deep learning approaches, particularly convolutional networks, have achieved significant success in various visual recognition

This study is partially supported by National Institutes of Health, National Institute of Biomedical Engineering and Bioengineering Award Number 5R01EB033413, National Institute of Diabetes and Digestive and Kidney Diseases Award Number 1R01124789, and by National Science Foundation (NSF) Center to Stream Healthcare in Place (#C2SHiP) CNS Award Number 2052578.

Abderrachid Hamrani is with Department of Mechanical and Materials Engineering, Florida International University, Miami, FL 33174, USA (e-mail: ahamrani@fiu.edu).
Daniela Leizaola is with Optical Imaging Laboratory, Department of Biomedical Engineering, Florida International University, Miami, FL 33174, USA (e-mail: dleiz001@fiu.edu).
Renato Sousa is with White Memorial Medical Group, Los Angeles, CA 90033, USA (e-mail: sousadpm@gmail.com).
Jose P. Ponce is with White Memorial Medical Group, Los Angeles, CA 90033, USA (e-mail: josepponce8@gmail.com).
Stanley Mathis is with White Memorial Medical Group, Los Angeles, CA 90033, and Clemente Clinical Research, Los Angeles, CA 90033, USA (e-mail: stan.mathis@clementeclinical.com).
David G. Armstrong is with Southwestern Academic Limb Salvage Alliance, Department of Surgery, Keck School of Medicine of University of Southern California, Los Angeles, CA 90033 (e-mail: armstrong.dg@gmail.com).
Anuradha Godavarty is with Optical Imaging Laboratory, Department of Biomedical Engineering, Florida International University, Miami, FL 33174, USA (e-mail: godavart@fiu.edu).

Color versions of one or more of the figures in this article are available online at http://ieeexplore.ieee.org

tasks, including biomedical image processing [13], [14]. Convolutional networks have evolved from basic architectures to more sophisticated designs, such as the U-Net, which was tailored for biomedical image segmentation [15], [16]. The U-Net architecture, with its contracting and expanding paths, has proven effective in capturing context and enabling precise localization. Building on these advancements, diffusion models introduce a fundamentally different approach to image synthesis and analysis, shifting the focus from deterministic mappings to probabilistic transformations. Diffusion models represent a groundbreaking advancement in the field of generative models, having gained significant attention for their ability to produce high-fidelity images that rival those generated by traditional methods [17]. Unlike earlier generative approaches such as generative adversarial networks (GANs) or variational autoencoders (VAEs) [18], [19], diffusion models operate through a unique process that incrementally adds noise to an image and then learns to reverse this process. This iterative denoising technique allows the model to gradually refine its predictions, enhancing its ability to reconstruct complex visual information from highly corrupted inputs. By systematically breaking down the image generation process into a series of reversible steps, these models gain a nuanced understanding of the data's underlying structure. This characteristic makes diffusion models particularly suited for tasks where detail and accuracy are paramount, such as medical imaging [20].

In the context of medical imaging, the strength of diffusion models lies in their exceptional ability to capture the subtle nuances and variations that are typical in medical data [21]. For instance, in imaging modalities like MRI or CT scans, slight variations in tissue density or abnormality size can be crucial for accurate diagnosis and treatment planning. Diffusion models' proficiency in handling such details offers a significant advantage over traditional segmentation methods, which often struggle with the variability and complexity of medical images. Moreover, a key challenge in applying advanced machine learning techniques to medical imaging has been the scarcity of high-quality, annotated datasets [22]. Annotating medical images requires expert knowledge and is time-consuming and costly. Diffusion models offer a promising solution to this bottleneck by reducing the reliance on extensive labeled datasets. Their ability to learn from and generate nuanced images in an unsupervised manner potentially alters how medical image segmentation tasks are approached [23], [24].

The objective of this study is to harness the capabilities of self-attention mechanisms within diffusion models to address the pressing need for effective, zero-shot image segmentation of diabetic foot ulcers. By doing so, this research aims to circumvent the limitations of data-dependency and improve the generalizability of segmentation models across varied clinical scenarios, thereby contributing to more accurate, efficient, and accessible diabetic foot ulcer management.

*A. Literature review*

The table below provides a summary of the methodologies used in various wound segmentation studies, emphasizing their datasets and performance metrics. The table emphasizes the evolution from traditional machine learning approaches to advanced deep learning models, showcasing the progression in segmentation accuracy across different datasets. By presenting a range of models, from MobileNet and U-Net to conditional GANs, this brief review outlines the state of the art in wound segmentation, particularly for DFUs and other chronic wounds.

TABLE I
LITERATURE REVIEW SUMMARY OF WOUND SEGMENTATION TECHNIQUES ACROSS DIFFERENT DATASETS.

| Ref | Methodology | Dataset |
| --- | --- | --- |
| [25] | WoundSeg using MobileNet and VGG16 | 950 images |
| [26] | MobileNetv2 with post-processing | 810 training, 200 test images |
| [27] | LinkNet and U-Net ensemble | FUSeg Challenge 2021 |
| [28] | FCN32 with VGG backbone | DFUC2022 (4000 images) |
| [29] | OCRNet with ConvNeXt | DFUC2022 |
| [30] | SegFormer MiT-B5 | DFUC2022 |
| [31] | K-means and SVM | 767 tissue regions |
| [32] | K-means and SVM | 96 images (64 training, 32 validation) |
| [33] | Attention-embedded encoder-decoder | Swift Medical Wound Data Set (467,000 images) |
| [33] | CNNs with morphological operations | Medetec database |
| [34] | Conditional GAN | eKare Inc. (100-4000 images) |
| [35] | Hybrid deep learning with mix Transformer | DFUTissue dataset (110 labeled, 600 unlabeled images) |
| [36] | Adaptive-gated MLP for location and image analysis | AZH and Medetec datasets |

The table underscores a significant gap in the literature, as most existing methods rely on supervised learning with large labeled datasets. In contrast, our approach ADZUS introduces a novel text-guided diffusion model that achieves competitive segmentation results without requiring labeled data.

*B. Motivation*

Producing high-quality segmentation masks for medical images remains a fundamental challenge in biomedical image analysis. Recent research has explored large-scale supervised training to enable zero-shot transfer segmentation and unsupervised methods to reduce reliance on dense annotations. However, constructing a model capable of segmenting diverse medical images in a zero-shot manner without any annotations remains difficult. Our motivation stems from the observation that stable diffusion models, initially designed for image generation, can produce highly realistic and detailed images (including medical ones) based solely on text prompts. Figure 1 illustrates this potential by comparing wounds generated using stable diffusion models with real clinical cases. The top row (a, b, c) displays diabetic foot ulcers generated using the following text prompt:

*"A highly detailed, realistic close-up of a diabetic foot ulcer. The open wound has irregular edges, with inflamed, reddish skin surrounding granulation tissue, necrotic (blackened) areas, and yellowish slough. The surrounding skin is dry, cracked, and discolored with mild swelling, indicating poor circulation. Captured in clinical lighting, the image highlights high-resolution textures of the skin and wound."*

The generated images exhibit notable resemblance to real clinical cases shown in the bottom row (d, e, f), capturing details such as granulation tissue, necrosis, and skin discoloration. This high similarity motivates the hypothesis that the self-attention mechanisms within stable diffusion models inherently capture medical imaging concepts, making them suitable for zero-shot segmentation tasks. By leveraging these properties, our approach aims to bypass the need for annotated datasets while maintaining high segmentation accuracy, thus offering a scalable solution for diverse medical imaging applications.

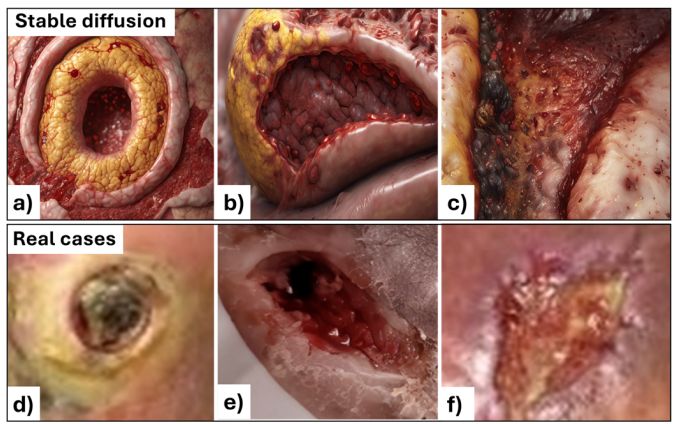

Fig. 1. Comparison of Stable Diffusion-generated (a, b, c) and real diabetic foot ulcer images (d, e, f).

Given this capability, an important question arises: has Stable Diffusion been trained on medical images (e.g., wound-related data), in particular the datasets used in this study? To address this, we conducted a data similarity verification analysis using the publicly available LAION-5B search tool [37], which forms the primary training dataset of Stable Diffusion. The images used in this study, including those in Figure 1 (d, e, f) and those in the results section, were tested to determine if they were present in Stable Diffusion's training corpus. Our analysis revealed no identical or closely matching images within the LAION-5B database. These findings confirm that our dataset was not included in the pretraining data of Stable Diffusion, ensuring that our segmentation experiments are conducted without prior model exposure to the images used in this study.

## II. METHODOLOGY

A pre-trained stable diffusion model is leveraged by ADZUS, with utilization of its self-attention layers to generate high-quality segmentation masks. In subsection II.A, a concise overview of the stable diffusion model architecture will be provided, followed by a detailed introduction to ADZUS in subsection II.B.

### A. Overview of the stable diffusion model

The Stable Diffusion Model [38], a well-known variant within the diffusion model family [39], [40], is a generative model that operates through both forward and reverse passes. During the forward pass, Gaussian noise is incrementally added at each time step until the image becomes completely isotropic Gaussian noise. Conversely, in the reverse pass, the model is trained to progressively eliminate this Gaussian noise, thereby reconstructing the original clean image. The stable diffusion model [38] incorporates an encoder-decoder and U-Net architecture with attention layers [39] (Figure 2).

Initially, an image $x \in \mathbb{R}^{H \times W \times 3}$ is compressed into a latent space with reduced spatial dimensions $z \in \mathbb{R}^{h \times w \times c}$ using an encoder $z = E(x)$. This latent space can then be decompressed back into the image $\tilde{x} = D(z)$ through a decoder. All diffusion processes occur within this latent space via the U-Net architecture, which is the primary focus of this paper's investigation. The U-Net consists of modular blocks, including 16 specific blocks composed of ResNet layers and Transformer layers. The Transformer layer uses two attention mechanisms: self-attention to learn global attention across the image and cross-attention to learn attention between the image and optional text input.

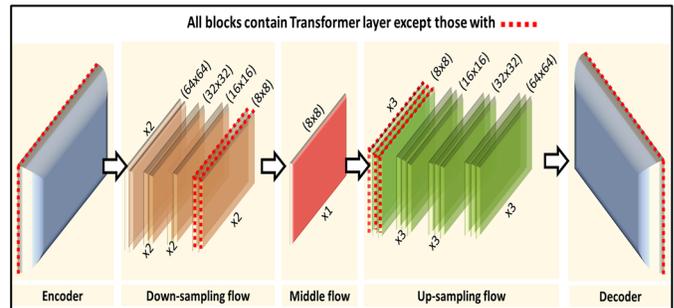

Fig. 2. Schematic of the stable diffusion configuration used in ADZUS model, consisting of 16 blocks, each containing transformer layers that produce a 4d self-attention tensor at various resolutions.

The component of interest for our investigation is the self-attention layer in the Transformer layer. Specifically, there are 16 self-attention layers distributed across the 16 composite blocks, resulting in 16 self-attention tensors. Each attention tensor $A_k \in \mathbb{R}^{h_k \times w_k \times h_k \times w_k}$ is 4-dimensional. Inspired by DiffuMask [41], which demonstrates object grouping in the cross-attention layer, it is hypothesized that the unconditional self-attention also contains inherent object grouping information, which can be used to produce segmentation masks without text inputs.

For each spatial location (*I*, *J*) in the attention tensor, the corresponding 2D attention map $A_k[I, J, :, :] \in \mathbb{R}^{h_k \times w_k}$ captures the semantic correlation between all locations and the location (*I*, *J*). Each location (*I*, *J*) corresponds to a region in the original image pixel space, the size of which depends on the receptive field of the tensor.

Two important observations motivate the method proposed in the next section:

- **Intra-Attention Similarity:** Within a 2D attention map $A_k[I,J,:,:]$, locations tend to have strong responses if they correspond to the same object group as (*I*, *J*) in the original image space.
- **Inter-Attention Similarity:** Between two 2D attention maps, e.g., $A_k[I,J,:,:]$ and $A_k[I+1,J+1,:,:]$, they tend to share similar activations if (*I*, *J*) and (*I* + *1*, *J* + *1*) belong to the same object group in the original image space.

The resolution of the attention map dictates the size of its receptive field concerning the original image. Lower resolution maps (e.g., 8×8) provide better grouping of large objects, while higher resolution maps (e.g., 16×16) offer more fine-grained grouping of components within larger objects, potentially identifying smaller objects more effectively. The current stable diffusion model has attention maps in four resolutions: 8×8, 16×16, 32×32, and 64×64. Building on these observations, a simple heuristic is proposed to aggregate weights from different resolutions and an iterative method to merge all attention maps into valid segmentation masks. In our experiments, the stable diffusion pre-trained models from "Huggingface" are used [42]. Typically, these prompt-conditioned diffusion models run for 50 or more diffusion steps to generate new images. However, to efficiently extract attention maps for an existing clean image without conditional prompts, we use only the unconditioned latent and run the diffusion process once. The unconditioned latent is calculated using an unconditioned text embedding. We set the time-step variable *t* to a large value (e.g., *t = 300*) so that real images are viewed as primarily denoised generated images from the diffusion model's perspective.

### B. ADZUS model

Since the self-attention layers capture inherent object grouping information in spatial attention (probability) maps, we propose ADZUS, a simple post-processing method, to aggregate and merge attention tensors into a valid segmentation mask. The pipeline consists of three components: attention aggregation, iterative attention merging, and non-maximum suppression. ADZUS is built on pre-trained stable diffusion models. For our implementation, we use stable diffusion V1.4 [38].

*1) Attention aggregation*

Given an input image passing through the encoder and U-Net, the stable diffusion model generates 16 attention tensors. Specifically, there are 5 tensors for each of the dimensions: (64 × 64 × 64 × 64), (32 × 32 × 32 × 32), (16 × 16 × 16 × 16), and (8 × 8 × 8 × 8). The goal is to aggregate attention tensors of different resolutions into the highest resolution tensor. To achieve this, the last 2 dimensions of all attention maps are up-sampled (bilinear interpolation) to 64 × 64, their highest resolution. Formally, for $A_k \in \mathbb{R}^{h_k \times w_k \times h_k \times w_k}$:

$$\tilde{A}_k = \text{Bilinear-upsample}(A_k) \in \mathbb{R}^{h_k \times w_k \times 64 \times 64} \quad (1)$$

The first 2 dimensions indicate the locations to which attention maps are referenced. Therefore, we aggregate attention maps accordingly. For example, the attention map in the (0, 0) location in $A_k \in \mathbb{R}^{8 \times 8}$ is first upsampled and then repeatedly aggregated pixel-wise with the 4 attention maps (0, 0), (0, 1), (1, 0), (1, 1) in $A_z \in \mathbb{R}^{16 \times 16}$. Formally, the final aggregated attention tensor $A_f \in \mathbb{R}^{64 \times 64}$ is:

$$A_f[I,J,:,:] = \sum_{k \in \{1,\ldots,16\}} \tilde{A}_k[I/\delta_k, J/\delta_k, :, :] * R_k \quad (2)$$

where $\delta_k = 64/w_k$ and $\sum_k R_k = 1$. The aggregated attention map is normalized to ensure it is a valid distribution. The weights *R* are important hyper-parameters and are proportional to the resolution $w_k$.

*2) Iterative attention merging*

In this step, the algorithm computes an attention tensor $A_f \in \mathbb{R}^{64 \times 64}$. The goal is to merge the 64×64 attention maps in the tensor $A_f$ to a stack of object proposals where each proposal likely contains the activation of a single object or category. Instead of using a K-means algorithm, which requires specifying the number of clusters, we generate a sampling grid from which the algorithm can iteratively merge attention maps. A set of $M \times M$ evenly spaced anchor points is generated. We then sample the corresponding attention maps from the tensor $A_f$. This operation yields a list of $M^2$ 2D attention maps as anchors:

$$L_a = \{A_f[im, jm, :, :] \in \mathbb{R}^{64 \times 64} | (im, jm) \in M\} \quad (3)$$

To measure similarity between attention maps, we use KL divergence:

$$2 * D(A_f[i,j], A_f[y,z]) = \\ (\text{KL}(A_f[i,j] \| A_f[y,z]) + \text{KL}(A_f[y,z] \| A_f[i,j])) \quad (4)$$

We start with N iterations of the merging process, where we compute the pair-wise distance between each element in the anchor list and all attention maps, averaging all attention maps with a distance smaller than a threshold $\tau$. This process is repeated in subsequent iterations, reducing the number of proposals by merging maps with distances smaller than $\tau$.

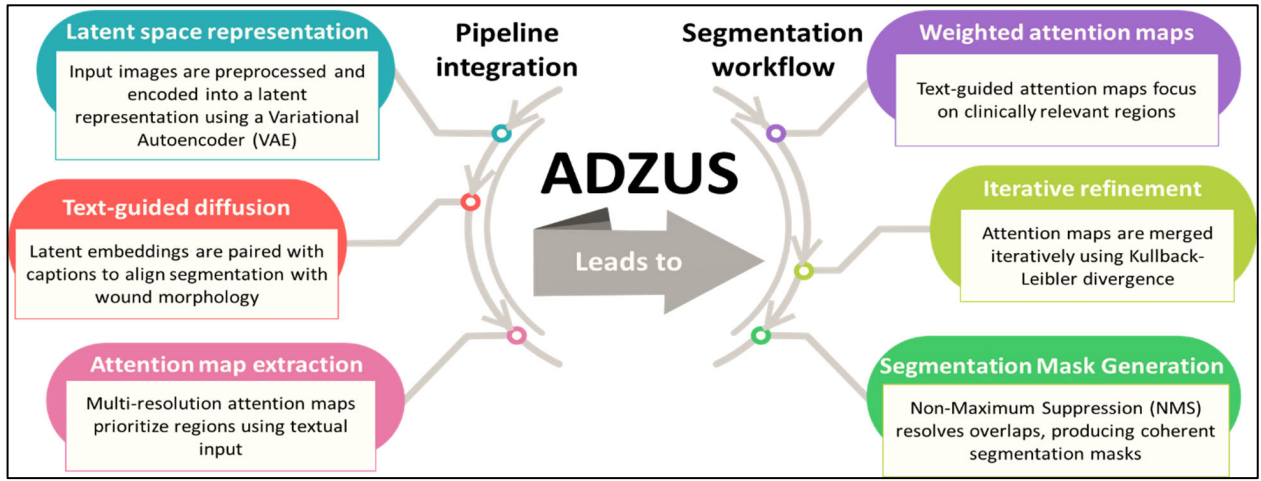

Fig. 3. Workflow of text-to-image diffusion integration in ADZUS for wound segmentation.

### 3) Non-maximum suppression

The iterative attention merging step yields a list $L_p \in \mathbb{R}^{N_p \times 64 \times 64}$ of $N_p$ object proposals in the form of attention maps. To convert the list into a valid segmentation mask, we use non-maximum suppression (NMS). Each element is a probability distribution map, and the final segmentation mask $S \in \mathbb{R}^{512 \times 512}$ is obtained by upsampling all elements in $L_p$ to the original resolution and taking the index of the largest probability at each spatial location across all maps. This methodology, combining attention aggregation, iterative merging, and non-maximum suppression, forms the core of the ADZUS approach for producing high-quality segmentation masks.

The iterative attention merging step yields a list of object proposals in the form of attention maps. To convert this list into a valid segmentation mask, we apply NMS, ensuring the selection of the most relevant segmented regions. Rather than directly identifying a specific anatomical structure, ADZUS generates a comprehensive segmentation mask that delineates multiple regions within the image, leveraging its self-attention mechanisms to outline the boundaries of all distinguishable structures. The model performs guided segmentation, generating a set of segmented regions without inherently classifying or labeling a specific area. Instead, it provides a structured segmentation output, allowing clinicians or users to interactively select the relevant region of interest based on the specific medical application.

### 4) Integration of text-to-image diffusion in ADZUS

The ADZUS framework utilizes the text-to-image capabilities inherent in stable diffusion models to enhance the segmentation process of diabetic wounds. This methodology enables the integration of descriptive textual inputs, thereby enriching the feature extraction process and improving the accuracy and robustness of the segmentation results.

A core novelty of the ADZUS model lies in its ability to incorporate detailed medical captions to guide the segmentation process. This integration enables the model to process latent embeddings of input images alongside descriptive textual prompts. The incorporation of textual guidance facilitates the model in capturing nuanced semantic features, particularly in scenarios where visual data alone may not sufficiently differentiate overlapping or ambiguous wound characteristics. This approach enhances segmentation precision by aligning generated attention maps with clinically relevant attributes.

The integration of text-to-image diffusion into the ADZUS framework follows a structured pipeline (Figure 3) that begins with preprocessing input images and encoding them into a compact latent representation using a Variational Autoencoder (VAE), which serves as the foundation for subsequent analysis. These latent embeddings are paired with descriptive captions to provide clinical context during the diffusion process, aligning segmentation outputs with detailed wound morphology features such as necrotic tissue, granulation areas, and fibrin deposits. Multi-resolution attention maps are then generated through self-attention mechanisms, prioritizing regions based on textual input. The segmentation workflow leverages these outputs through three key steps: generating text-guided weighted attention maps, iteratively refining and merging attention maps using Kullback-Leibler divergence and creating final segmentation masks via Non-Maximum Suppression (NMS) to resolve overlaps and produce coherent, clinically aligned outputs.

## III. RESULTS

The results presented in this study aim to evaluate the performance of ADZUS on a custom-curated dataset of wound images, focusing on its ability to accurately segment wound regions. The analysis compares ADZUS against conventional methods, including CNN and K-means, as well as state-of-the-art segmentation models such as Swin-Unet and DeepLabV3Plus [43], [44]. Additionally, the concept of text-guided image segmentation, demonstrating how descriptive prompts can dynamically influence the segmentation process, is introduced. This study evaluated ADZUS's effectiveness in wound tissue segmentation using established metrics, including dice similarity coefficient (DSC), intersection-over-union (IoU), precision, and recall. A detailed description of these 6metrics is provided in the Appendix A.

## A. Comparison ADZUS with state-of-the-art methods

The first results provide a comparative analysis of ADZUS against state-of-the-art segmentation models using the publicly available chronic wound dataset [26]. This dataset consists of 1,010 labeled images of diabetic foot ulcers, with 810 images designated for training and 200 for inference. It provides a standardized platform for segmentation performance assessment within a supervised learning framework. The state-of-the-art models includes LinkNet-EffB1+UNet-EffB2 [27], [43], DeepLabV3Plus [44], [45], Swin-Unet [43], DDRNet [43], SegFormer-b5 [46], and FUSegNet [43], which were used for the comparison.

Having ADZUS as an unsupervised learning model presents a significant challenge in this evaluation since supervised models leverage labeled wound tracings to optimize their performance during training. The experiments detailed in this section aim to contextualize ADZUS's performance within the broader landscape of advanced segmentation techniques, emphasizing its potential to deliver accurate and adaptable results despite its independence from labeled training data. In this evaluation, the prompt text used for ADZUS was: "*a detailed medical photograph of a diabetic foot ulcer with necrotic tissue, slough formation, granulation areas, and wound exudate surrounding tissue damage.*" This descriptive input guided the segmentation process, allowing ADZUS to generate meaningful delineations of wound structures without relying on labeled supervision.

As illustrated in Table II, ADZUS demonstrates competitive performance (using only the test dataset of 200 images) when compared with state-of-the-art segmentation models. Notably, ADZUS outperforms the previously best-performing FUSegNet model by achieving the highest IoU of 86.68% and the highest precision of 94.69%, compared to FUSegNet's IoU of 86.40% and precision of 94.40%. In terms of recall score, ADZUS achieves an impressive 92.46%, exceeding most models, including DeepLabV3Plus and LinkNet-EffB1 + UNet-EffB2, while remaining slightly behind in DSC compared to FUSegNet, which achieved the highest DSC of 92.70%. These findings collectively highlight the strength of ADZUS in delivering accurate and reliable segmentation results without the need for labeled training data, a significant advantage over supervised models such as FUSegNet and SegFormer-b5.

TABLE II
PERFORMANCE COMPARISON OF ADZUS AND STATE-OF-THE-ART MODELS BASED ON EVALUATION METRICS.

| Model | IoU (%) | Precision (%) | Recall (%) | DSC (%) |
|---|---|---|---|---|
| DDRNet | 57.64 | 80.86 | 66.75 | 73.13 |
| Swin-Unet | 79.30 | 89.94 | 87.02 | 88.46 |
| SegFormer-b5 | 83.58 | 92.21 | 89.94 | 91.06 |
| DeepLabV3Plus | 85.19 | 92.75 | 91.27 | 92.00 |
| LinkNet-EffB1 + UNet-EffB2 | 85.51 | 92.68 | 91.80 | 92.07 |
| FUSegNet | 86.40 | 94.40 | 91.07 | **92.70** |
| ADZUS (our model) | **86.68** | **94.69** | **92.46** | 91.98 |

The qualitative results for the chronic wound dataset are illustrated in Figure 4. The segmentation outputs demonstrate the performance of ADZUS in accurately delineating wound regions, with original boundaries depicted in red and predicted boundaries in green. The images are cropped to enhance visualization and focus on the wound areas. Another qualitative results for the chronic wound dataset are presented in Figure 5, offering a comparative analysis of segmentation outputs from ADZUS and benchmark models. Original boundaries, shown in red, and predicted boundaries, depicted in green, are overlaid on cropped images to enhance clarity and focus on the wound regions. The figure highlights ADZUS's ability to achieve precise segmentation, closely aligning with the original boundaries and minimizing deviations. With a DSC of 93.56, ADZUS demonstrates competitive performance, surpassing supervised models like FUSegNet and SegFormer, while performing slightly below DeepLabV3+. This analysis underscores the effectiveness of ADZUS's text-guided diffusion mechanism and self-attention features in accurately delineating wound boundaries, even in complex and challenging scenarios.

## B. Comparison ADZUS with FUSegNet using a custom-curated dataset

Building upon the promising results from the comparison with state-of-the-art segmentation models, this section evaluates ADZUS against FUSegNet model using a custom-curated dataset of 40 white-light (WL) images of diabetic foot ulcers (DFUs) with ground truth tracings of the wounds (by the clinical team). The WL images, were captured using a smartphone-based near-infrared optical imaging device [47], [48] from 15 participants over 1-4 weeks (in an IRB approved study), were manually traced by clinicians to delineate wound regions. This curated dataset ensures a consistent basis for comparing the performance of both methods under identical conditions. Unlike ADZUS, which requires no labeled training data, FUSegNet [43] is a supervised model trained on the FUSeg dataset, which consists of 1,210 foot ulcer images, including 1,010 images for training and 200 images for evaluation. The previously used prompt text for ADZUS has been maintained for this comparison.

Quantitative analysis (Figure 6) demonstrated ADZUS's superior performance over FuSegNet, with a median DSC score of approximately 75% and an IoU score of around 68%. In contrast, FuSegNet exhibited lower segmentation accuracy, with median DSC and IoU scores of approximately 45% and 50%, respectively. Figure 6 highlights these differences, showing that ADZUS consistently achieved higher segmentation agreement with ground truth across varying conditions.

Qualitative comparisons further reinforce these findings. Figure 7 shows that ADZUS effectively segmented wound boundaries, avoiding over-segmentation and peri-wound misclassification, which were common issues with FUSegNet. The superior performance of ADZUS is attributed to its text-guided diffusion mechanism and self-attention capabilities, which enabled precise identification of granulation zones while excluding irrelevant features.

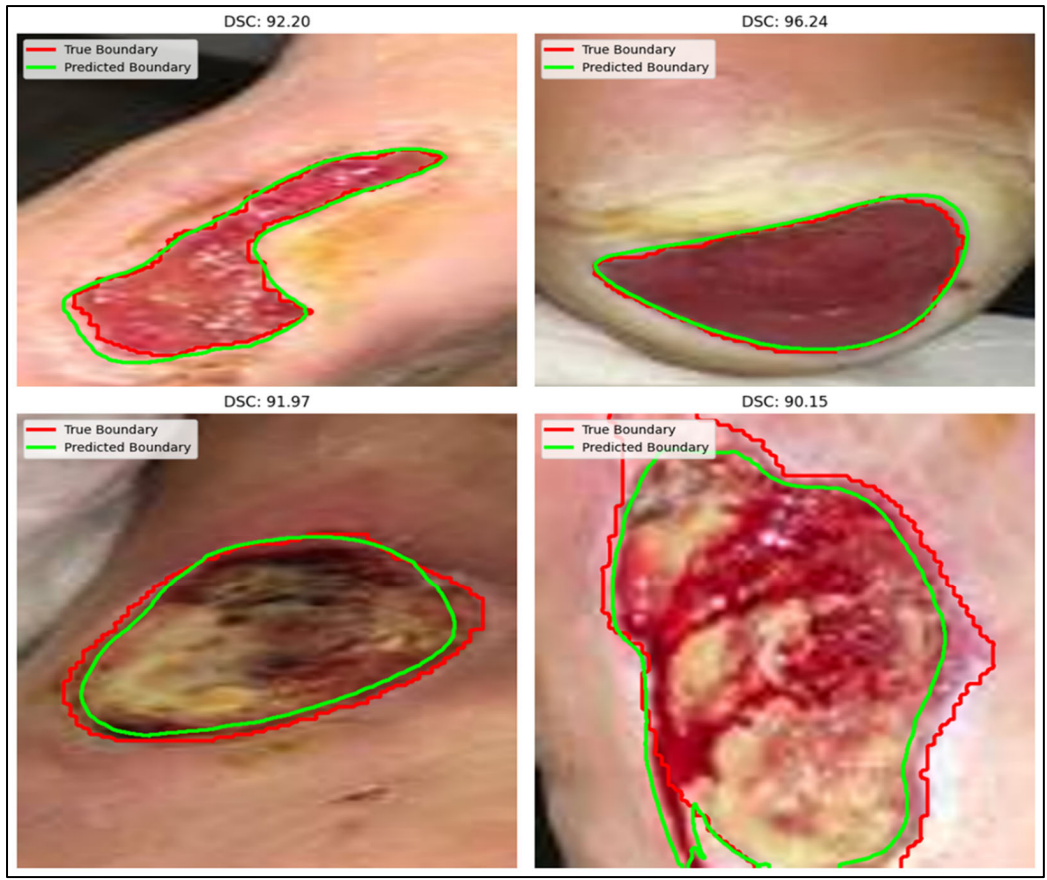

Fig. 4. Sample qualitative segmentation on the chronic wound dataset [26]: original boundaries (red) and predicted boundaries (green) displayed on cropped images for enhanced visualization.

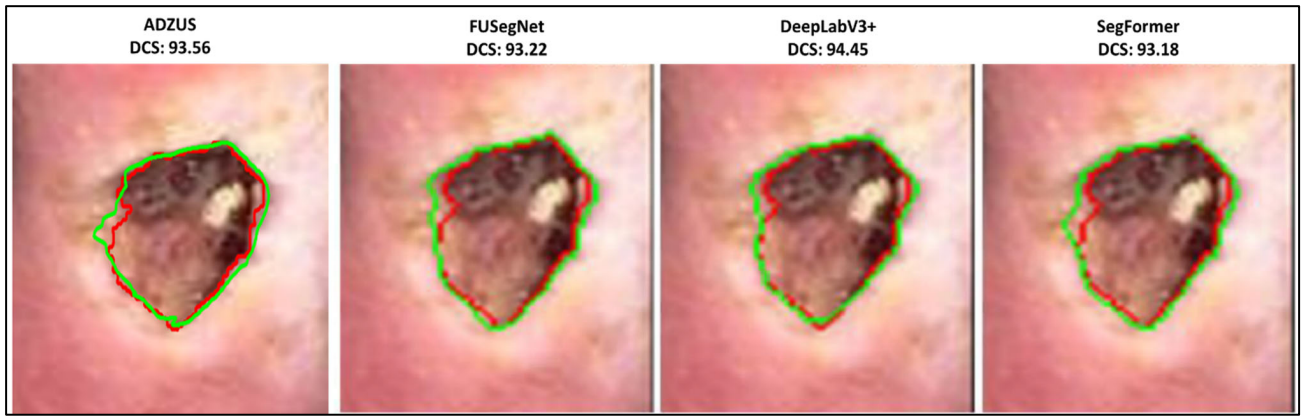

Fig. 5. Sample comparison of segmentation results: ADZUS vs. benchmark models on chronic wound dataset [26], showing original boundaries (red) and predicted boundaries (green) on cropped images.

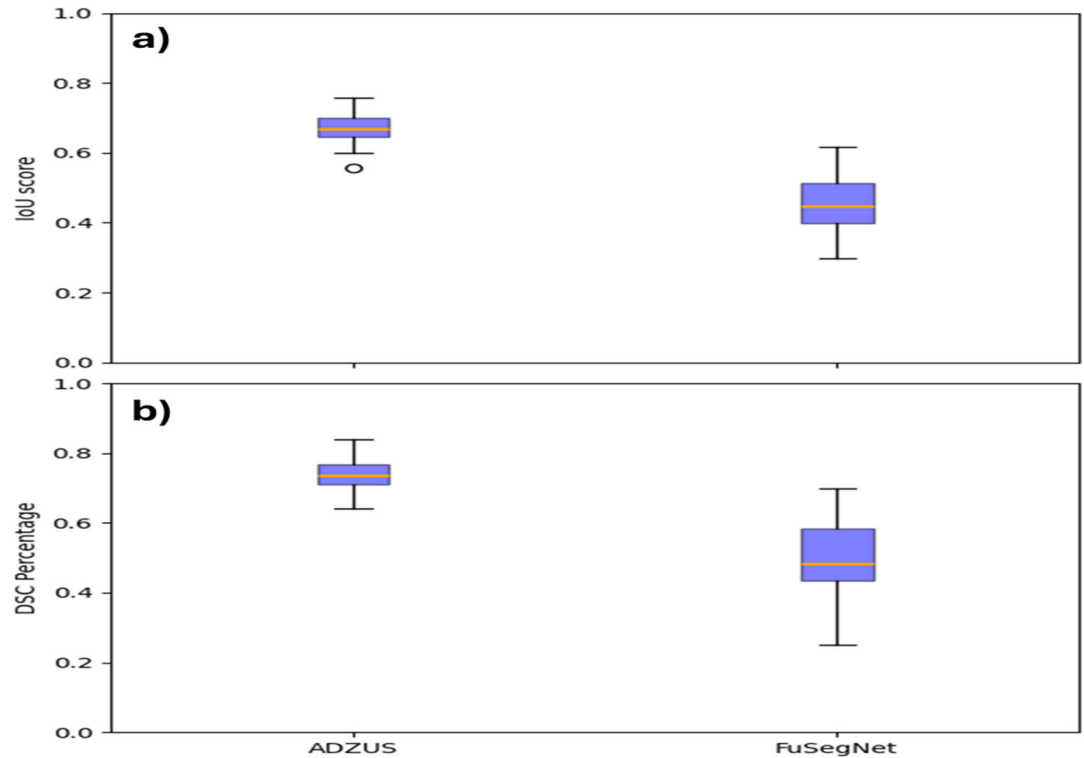

Fig. 6. Comparison of ADZUS with FUSegNet model based on (a) IoU and (b) DSC scores.

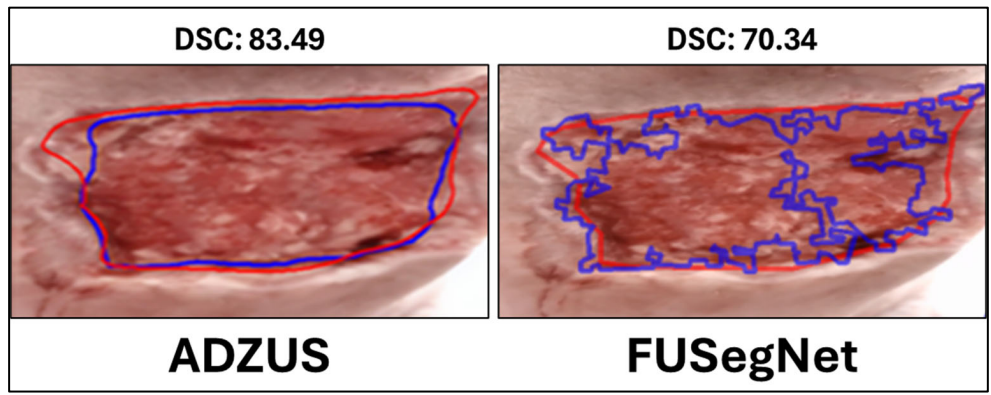

Fig. 7. Sample comparison of wound segmentation results using ADZUS and FUSegNet for one sample DFU case from our custom-curated dataset. Red outlines represent clinical ground truth tracings, while blue outlines denote segmentations produced by the respective methods.

C. Demonstration of text-guided segmentation

The ability of ADZUS to perform text-guided segmentation is demonstrated in Figure 8, where different descriptive prompts were used to influence the segmentation results (using an image from our custom-curated dataset of 40 white-light (WL) images of DFUs). The leftmost panel presents the original wound image, while the middle and right panels illustrate segmentation outputs generated based on two distinct textual prompts. The segmentation maps incorporate confidence scores, visualized through a color scale, with higher confidence regions indicated by red and lower confidence by blue. In the first scenario, the descriptive prompt "a detailed medical photograph highlighting infected wound" leads to a segmentation that focuses on broader wound structures, identifying general ulcer boundaries with relatively uniform confidence across the segmented regions. In contrast, a second prompt, "a wound with inflamed red tissue, swollen areas, and early signs of infection", enhances the segmentation granularity, highlighting additional pathological features such as peri-wound inflammation and early-stage infection indicators. This differentiation underscores ADZUS's adaptability in generating hierarchical segmentations based on varying levels of textual detail. The first segmentation output primarily captures the overall wound region, while the second introduces finer delineations of tissue abnormalities. The accuracy of this text-guided segmentation, correlating the text prompts provided and regions it segmented, is part of our ongoing efforts.

By integrating a text-guided diffusion mechanism, ADZUS enables customized segmentation that aligns with specific clinical descriptions, providing a flexible and interactive tool for medical image analysis.

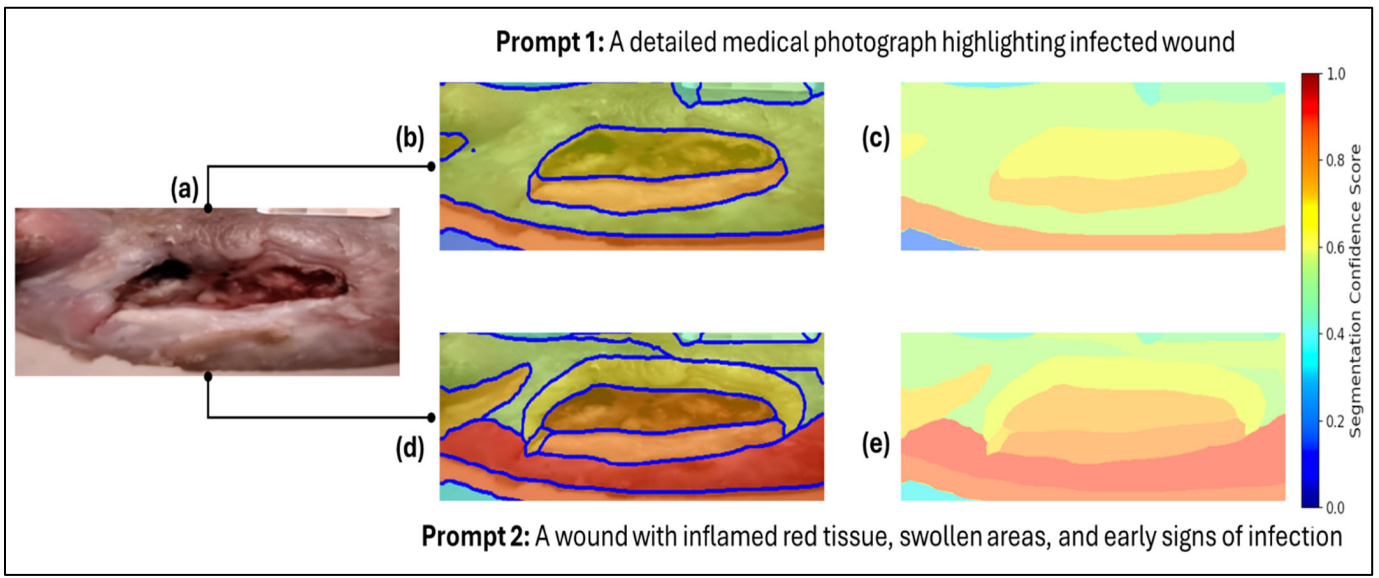

Fig. 8. Demonstration of text-guided segmentation in ADZUS: (a) original wound image, (b) segmentation result using Prompt 1, (c) confidence map corresponding to Prompt 1, (d) segmentation result using Prompt 2, and (e) confidence map corresponding to Prompt 2.

## IV. Discussion

The findings of this study demonstrate the effectiveness of ADZUS in wound segmentation, highlighting its ability to perform competitively against state-of-the-art supervised models while operating in a zero-shot learning mode. Unlike traditional approaches that rely on extensive labeled datasets, ADZUS segments wound regions without prior training, leveraging its text-guided diffusion mechanism for precise boundary delineation. The comparison with established segmentation models, including FUSegNet, DeepLabV3Plus, and Swin-Unet, underscores its robustness. On the publicly available chronic wound dataset, ADZUS achieved the highest IoU (86.68%) and precision (94.69%) among all evaluated models, surpassing FUSegNet, which had previously been the best-performing approach. While ADZUS exhibited a slightly lower DSC (91.98%) than FUSegNet (92.70%), its superior recall indicates a strong ability to capture wound regions effectively, reducing under-segmentation risks and ensuring comprehensive tissue identification.

Further evaluation on a custom-curated dataset of white-light images of diabetic foot ulcers reaffirmed ADZUS's superiority over FUSegNet, achieving a median DSC of 75% and an IoU of 68%, significantly outperforming FUSegNet, which reached DSC and IoU scores of around 45% and 50%, respectively. Qualitative analysis reinforced these findings, with ADZUS demonstrating greater consistency in delineating wound boundaries while mitigating peri-wound misclassification.

One of the key contributions of this study is the demonstration of ADZUS's text-guided segmentation capability, allowing dynamic refinement of segmentation outputs based on descriptive textual prompts. By adjusting segmentation boundaries according to specified wound characteristics, ADZUS introduces a level of flexibility not present in traditional models. The ability to incorporate user-defined textual input enables more precise segmentation tailored to specific clinical needs, as demonstrated by the differentiation of necrotic tissue, granulation zones, and exudate accumulation based on distinct textual descriptions. This feature allows ADZUS to refine segmentation results based on prompt-specific details, where different textual descriptions led to the segmentation of distinct wound zones, providing an interpretable and adaptable segmentation approach that aligns closely with expert clinical assessment.

Despite these promising results, certain limitations must be addressed to enhance the model's practical applicability. The diffusion-based inference process introduces high computational costs and prolonged processing times, which could limit scalability in real-time clinical environments. Future work should focus on optimizing ADZUS's computational efficiency through reduced-step diffusion processes or hardware acceleration strategies. Additionally, while the model performs well across wound segmentation tasks, expanding its evaluation to broader medical imaging applications, such as retinal vessel segmentation or tumor boundary detection, could further validate its generalizability. The integration of domain adaptation techniques and hybrid self-supervised learning approaches may enhance performance in challenging segmentation scenarios. Furthermore, leveraging its text-guided segmentation capabilities in multimodal clinical applications, such as incorporating electronic health records or histopathological data, could unlock new possibilities for AI-driven medical diagnostics.

## V. Conclusion

In this study, we introduced the attention diffusion zero-shot unsupervised system (ADZUS) as a novel approach for wound segmentation, specifically targeting DFUs. The results demonstrated the model's ability to achieve precise segmentation without the reliance on labeled training data, distinguishing it from conventional and state-of-the-art supervised methods. Through a series of comparative analyses, ADZUS consistently achieved competitive performance across key evaluation metrics such as IoU, precision, recall, and DSC. Notably, the model surpassed several benchmark models, emphasizing its potential to deliver accurate wound boundary delineations with minimal data dependency.

The key innovation of ADZUS lies in its text-guided diffusion mechanism, which enables dynamic segmentation outputs tailored to specific descriptive prompts. This feature showcases the model's adaptability, allowing for customized wound analysis based on clinical descriptions. However, the study also highlighted certain limitations. The computational time required for text-guided segmentation remains a challenge. Furthermore, while the zero-shot approach offers flexibility, fine-tuning the model to accommodate diverse clinical datasets and contexts is necessary to enhance its robustness across various medical applications.

Looking ahead, future research should focus on optimizing the computational efficiency of ADZUS to facilitate its deployment in real-world clinical environments. Additionally, integrating multimodal data sources, such as electronic health records and advanced imaging modalities, could further enrich the model's predictive capabilities and broaden its utility in medical diagnostics. Expanding the model's validation across larger, more diverse datasets will be essential to ensure its generalizability and reliability in different healthcare contexts. Future research will focus on expanding ADZUS's capabilities to additional medical imaging domains, exploring integration with real-time diagnostic systems, and advancing toward a more autonomous segmentation framework. This next phase will enable ADZUS to recognize, label and segment specific zones of interest requested by clinicians, such as infections or abnormal regions, leveraging its learned attention patterns. The promising results of this study suggest that ADZUS could evolve into an intelligent, semi-supervised system capable of automatic region identification and preliminary labeling, thereby enhancing its adaptability for clinical applications while maintaining interpretability and expert oversight.

## APPENDIX

### Evaluation metrics

In this study, the evaluation of segmentation performance was critical to understanding the effectiveness of the ADZUS model for medical image segmentation. Established metrics (expressed as percentages), including dice similarity coefficient (DSC), intersection-over-union (IoU), precision, and recall, were adopted.

$$DSC = \left(\frac{2TP}{2TP + FP + FN}\right) \times 100 \text{,} \quad (A1)$$

$$IoU = \left(\frac{TP}{TP + FP + FN}\right) \times 100 \text{,} \quad (A2)$$

$$precision = \left(\frac{TP}{TP + FP}\right) \times 100 \text{,} \quad (A3)$$

$$recall = \left(\frac{TP}{TP + FN}\right) \times 100 \text{,} \quad (A4)$$

Where TP, FP, and FN represent true positives, false positives, and false negatives, respectively.

ACKNOWLEDGMENT

We would like to thank the residents, coordinators, and medical staff at Clemente Clinical Research and White Memorial Medical Group in LA for assisting us during clinical imaging studies.